\documentclass{article}

\PassOptionsToPackage{numbers, compress}{natbib}
\usepackage[final]{nips_2017}
\usepackage[utf8]{inputenc} 
\usepackage[T1]{fontenc}    
\usepackage{hyperref}       
\usepackage{url}            
\usepackage{booktabs}       
\usepackage{amsfonts}       
\usepackage{nicefrac}       
\usepackage{microtype}      
\usepackage{graphicx}
\usepackage{subcaption}
\usepackage{amsmath}

\title{Chest X-ray Inpainting with Deep Generative Models}

\author{
  \begin{tabular}{cc}
  Ecem Sogancioglu\thanks{Equal contribution.} & Shi Hu\textsuperscript{*} \\
  \normalfont Radboud University Medical Center & \normalfont University of Amsterdam \\
  \texttt{ecem.lago@radboudumc.nl} & \texttt{s.hu@uva.nl} \\
  \\
  Davide Belli & Bram van Ginneken \\ 
  \normalfont University of Amsterdam & \normalfont Radboud University Medical Center \\  
  \texttt{davidebelli95@gmail.com} & \texttt{bram.vanginneken@radboudumc.nl} \\
\end{tabular}
}

\begin{document}

\maketitle

\begin{abstract}

Generative adversarial networks have been successfully applied to inpainting in natural images. However, the current state-of-the-art models have not yet been widely adopted in the medical imaging domain. In this paper, we investigate the performance of three recently published deep learning based inpainting models: context encoders, semantic image inpainting, and the contextual attention model, applied to chest x-rays, as the chest exam is the most commonly performed radiological procedure. We train these generative models on 1.2M 128 $\times$ 128 patches from 60K healthy x-rays, and learn to predict the center 64 $\times$ 64 region in each patch. We test the models on both the healthy and abnormal radiographs. We evaluate the results by visual inspection and comparing the PSNR scores. The outputs of the models are in most cases highly realistic. We show that the methods have potential to enhance and detect abnormalities. In addition, we perform a 2AFC observer study and show that an experienced human observer performs poorly in detecting inpainted regions, particularly those generated by the contextual attention model. 



\end{abstract}

\section{Introduction}


Among different technologies for diagnosing and screening chest abnormalities, x-rays are the most widely employed examination due to its low cost and low radiation dose. In the United States around 150 million chest radiography exams are performed every year, and worldwide this number is around 10 times larger \cite{Mett09}. Abnormalities may be missed by readers due to high workload, fatigue, overprojection of normal structures, and interpretation errors. These problems call for an automated system that can assist radiologists in the accurate detection of abnormal regions. To accomplish this goal, one approach is to a supervised system to directly classify regions in the x-ray as normal or suspicious. Alternatively, one can use generative models to generate what a region should look like given its surrounding and the assumption it is normal, and then compare this generated region with the original one and measure their differences. In this work, we adopt the second approach for abnormality detection. This approach has the advantage that only normal images would be needed to build the generator; no annotations and abnormal images would be required. The main challenge of this approach is to generate a realistic patch that not only looks sensible on its own, but also fits well within the context of the rest of the image. To the best of our knowledge, this is the first work to focus specifically on generating realistic looking medical image patches.

Since their introduction in 2014, generative adversarial nets (GANs) \cite{gan} have shown great promise in generating realistic natural images and they have been improved greatly since then. Recently, the methodology of GANs, based on adversarial training, has been used and adapted to medical images for several purposes such as image synthesis \cite{histo}, denoising \cite{wolter} or segmentation \cite{scan}. 


Another area which took advantage of the GAN framework is semantic image inpainting, where the task is to generate (inpaint) a missing region in an image, conditioned on the context of the rest of the image. In order to succeed at this task, models have to recover not only the textured pattern but also understand the scene semantically. Recently, several works investigated GANs for this task \cite{semantic,contenc,gc,yu} and showed promising results for natural images. However, to the best of our knowledge, their applicability to medical images has not been investigated yet. 

Inpainting on medical images is very challenging since an erroneously inpainted region might completely change the meaning of the scene. For example, for chest x-rays, generating a round dense region in an otherwise normal image might lead to the incorrect conclusion that the patient may be at risk for lung cancer. On the other hand, if a model could generate a normal looking region for a given region that contains an abnormality, subtracting this generated region from the original images will make the abnormality stand out more clearly. Another application could be to remove foreign objects from chest radiographs. Such objects (pacemakers, catheters, drainage tubes) are commonly seen on chest x-rays. 

\section{Related Work}

Since this work in at the crossing of machine learning and medical image analysis, we provide two orthogonal sections of literature reviews. First we discuss general inpainting methodologies, and subsequently we focus on applications of inpainting in medical imaging.

Image inpainting is a well studied area which was covered by a large body of literature and can be divided into two main subgroups: classical inpainting methods and learning based methods. Classical approaches involve local and non-local approaches. Local approaches such as \cite{Bertalmio} use the information in the available part of the image whereas non local approaches attempt to solve this task by utilizing external data. Examples of local methods are given in \cite{texturesyn}, where one searches for a similar texture from the same image by employing non parametric sampling or a popular PatchMatch \cite{patchmatch}. These methods assume that the required patch to be inpainted can already be found elsewhere in the input image \cite{semantic}, which might work well to inpaint small textured regions, but will fail to convincingly complete larger regions with structural information. To overcome this limitation, as an example of non local approach, \cite{Hays} proposed an approach which uses a database of millions of images. Basically, the algorithm searches for an image in a huge dataset which is the most similar to the input image, and it inpaints the holes of the input image with the corresponding region of the matched image. However, this approach requires a huge dataset which is very hard to satisfy in the medical domain. Moreover, it also assumes that exact same scene will be in the dataset, which limits the applicability of the approach.

Several previous works in the medical imaging community have intersections with this work. Hogeweg et al.~proposed a method to automatically detect, segment, and remove foreign objects from chest radiographs \cite{Hogeweg}. They used a kNN classifier to perform pixel level classification and filled the detected regions using texture synthesis, whereas we use deep learning based models which perform better in content restoration. Litjens et al.~proposed a method to simulate nodules and diffuse infiltrates in chest radiographs \cite{Litjens}. They used projected and CT data to generate patches of nodules and infiltrates, and apply rotation and scaling for post-processing. These patches are then blended in the x-rays. This technique can not generate new normal image structures, such as ribs, and thus cannot be used for inpainting. In terms of using generative models to detect abnormalities in medical images, Schlegl et al.'s work \cite{anogan} is similar to ours. However, their paper largely focuses on classification and does not perform inpainting, whereas our work has a focus on the quality of the generated inpainted areas.


\section{Data}

We use the publicly available ChestXray14 dataset \cite{chestxray8}, which contains 112,120 frontal chest x-rays from 30,805 unique patients. All x-rays have been downsampled to 8-bit grayscale with dimension 1024 $\times$ 1024, and the presence or absence of 14 different thoracic abnormalities is indicated. The labels were extracted from the corresponding radiology reports using natural language processing. 

Since our aim is not only to inpaint the missing region with semantically and visually plausible pixels, but also fill this area with a healthy looking pattern, our generative models are trained on a subset of images from the ChestXray14 data set, namely those for which no abnormalities were indicated. We randomly selected 59,481 chest x-rays for training. For the test set, we selected 880 random cases among x-rays labeled with at least one abnormality and 880 random cases among normal x-rays. We made sure that no images from patients in the training set were included in the test set.  

\section{Methods}

We investigate three recently proposed inpainting methods, namely, the context encoders \cite{contenc}, the semantic inpainting model \cite{semantic} and the contextual attention model \cite{yu}.

\subsection{Context Encoders}
The context encoder trains a convolutional neural network to generate the contents of an arbitrary image region conditioned on its surroundings \cite{contenc}. Specifically, we feed the input image with a mask to an autoencoder, and it outputs the generated content under the mask. Both the encoder and the decoder in the autoencoder are convolutional networks, where the encoder has six layers and the decoder has five. Furthermore, the discriminator is also a convolutional network with five layers. The last layer outputs a scalar value between 0 and 1 indicating how real its input is. 

The model applies two loss functions to the output of the autoencoder, one is the pixel-wise reconstruction loss, and the other is the adversarial loss. This combined loss produces sharp and semantically meaningful results, and the loss has the form $L = \lambda_{rec}L_{rec} + \lambda_{adv}L_{adv}$. We assign $\lambda_{rec} = 0.998$ for the reconstruction loss and $\lambda_{rec} = 0.002$ for the adversarial loss. Our implementation is based on the publicly available PyTorch \cite{pytorch} code \footnote{\url{https://github.com/BoyuanJiang/context_encoder_pytorch}}, but we modify their code by doubling the initial channels from 64 to 128 for both the autoencoder and the discriminator. This modification results in better image quality.

\subsection{Semantic Image Inpainting} 

The semantic image inpainting \cite{semantic} model proposed by Yeh et al.~is trained in two steps. First, we train a deep generative model DCGAN \cite{Radford} on real images. After this generative training, we obtain a mapping between a latent code and its corresponding image produced by DCGAN. Then we fix the weights in DCGAN, and iteratively find the latent code $z$ whose mapped image matches the best with the original image outside a mask under a weighted context loss. This means, the generated pixels closer to the missing patch need to match more closely to the original image than distant pixels. In addition, they have a prior loss which penalizes unrealistic images. The loss is different than the context loss as the prior loss uses GAN to discriminate high-level image features; whereas the context loss is a weighted $l1$ loss. Our implementation is based on the TensorFlow \cite{tensorflow} implementation from the blog written by Brandon Amos \cite{amos2016image}, but we modified the code to have the exact implementation of the original paper.


\subsection{Contextual Attention Model}
Yu et al.~recently proposed the contextual attention model that is able to synthesize novel image structures as well as blend in textures from surrounding regions \cite{yu}. The training of this model consists of two stages, which goes through the coarse-to-fine transition in the quality of the generated patch. In the first stage, a dilated convolutional network is used to produce an initial coarse prediction. In the second stage, the coarse prediction from the first stage becomes the input and it is split into 2 paths. One is through another dilated convolutional network, the other is through a contextual attention layer. The results from these two paths are concatenated together and fed to another neural network to predict refined results. This model was shown to improve upon the method \cite{gc} for inpainting in natural images. We use the TensorFlow implementation provided by the authors \footnote{\url{https://github.com/JiahuiYu/generative_inpainting}}.

\section{Experiments}

\subsection{Setup}

To obtain the input training images for the generative models, we first use a popular medical image segmentation tool U-Net \cite{unet} to outline the lung locations. Then we randomly crop out 20 image patches of size 128 $\times$ 128 that have an overlap with the lung fields from each chest x-ray. This provides 1.2M patches for training. For each image patch, we mask out the center 64 $\times$ 64 patch and let our model to generate it. 

Training was performed on a variety of modern GPUs in a cluster, using one GPU per model. The GPU models we used include Nvidia 1080 Ti, Nvidia Titan Xp and Nvidia Titan V. Training took approximately 5 days for context encoders to reach 70 epochs, 6 days for semantic image inpainting to reach 80 epochs, and 7 days for the contextual attention model for 55 epochs.    

\subsection{Results}


Figure 1 displays five chest x-ray patches randomly chosen from the 880 \textit{healthy} ones in the test set. We compare the generative results of the three models. The right most column shows the original five patches, and the left most column is the input patch to the algorithms, with the masked out region to be inpainted in the center. Ideally, the generative models would produce a result identical to the original patches. The middle three columns are the results from the three generative models, namely, the context encoder, the semantic inpainting model, and the contextual attention model.

In general, we find that the context encoder is able to generate sharp and semantically meaningful patches; however, there are some discontinuities at the borders of the generated patch. On the other hand, the contextual attention model has better global consistency and therefore the borders of the inpainted region are often hardly noticeable. However, the content inside the patch tends to be a bit blurrier than the context encoder. The semantic inpainting model can sometimes produce good results, but its performance in general falls short to the other two. Nevertheless, all models can produce realistic looking results. Table \ref{psnr-healthy-compares} provides the means and standard deviations of the peak signal-to-noise ratio (PSNR) over all 880 healthy images in the test set. The PSNR value is computed with respect to the center 64 $\times$ 64 missing patch rather than the entire patch. We note the semantic inpainting model achieves the highest average PSNR in this case; however, this quantitative assessment doesn't fully correlate with the visual results.

\begin{figure}[t]
\centering
\begin{subfigure}[t]{0.18\textwidth}
\raisebox{-\height}{\includegraphics[width=\textwidth]{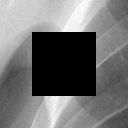}}
\raisebox{-\height}{\includegraphics[width=\textwidth]{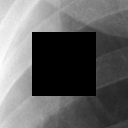}}
\raisebox{-\height}{\includegraphics[width=\textwidth]{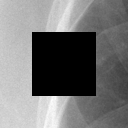}}
\raisebox{-\height}{\includegraphics[width=\textwidth]{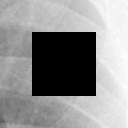}}
\raisebox{-\height}{\includegraphics[width=\textwidth]{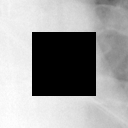}}

\caption{Input}
\end{subfigure}
\begin{subfigure}[t]{0.18\textwidth}
	\raisebox{-\height}{\includegraphics[width=\textwidth]{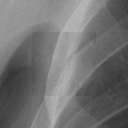}}
	\raisebox{-\height}{\includegraphics[width=\textwidth]{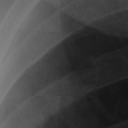}}
    \raisebox{-\height}{\includegraphics[width=\textwidth]{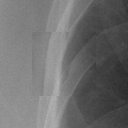}}
     \raisebox{-\height}{\includegraphics[width=\textwidth]{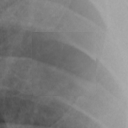}}
     \raisebox{-\height}{\includegraphics[width=\textwidth]{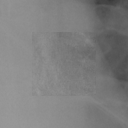}}

\caption{CE}
\end{subfigure}
\begin{subfigure}[t]{0.18\textwidth}
\raisebox{-\height}{\includegraphics[width=\textwidth]{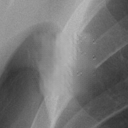}}
\raisebox{-\height}{\includegraphics[width=\textwidth]{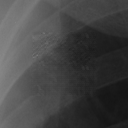}}
\raisebox{-\height}{\includegraphics[width=\textwidth]{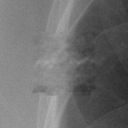}}
 \raisebox{-\height}{\includegraphics[width=\textwidth]{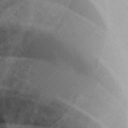}}
 \raisebox{-\height}{\includegraphics[width=\textwidth]{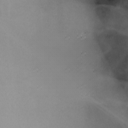}}

\caption{SI}
\end{subfigure}
\begin{subfigure}[t]{0.18\textwidth}   
\raisebox{-\height}{\includegraphics[width=\textwidth]{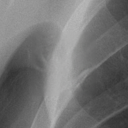}}
\raisebox{-\height}{\includegraphics[width=\textwidth]{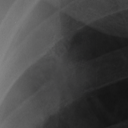}}
\raisebox{-\height}{\includegraphics[width=\textwidth]{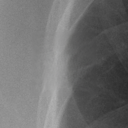}}
 \raisebox{-\height}{\includegraphics[width=\textwidth]{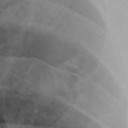}}
 \raisebox{-\height}{\includegraphics[width=\textwidth]{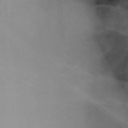}}
 
\caption{CA}
\end{subfigure}
\begin{subfigure}[t]{0.18\textwidth}   
\raisebox{-\height}{\includegraphics[width=\textwidth]{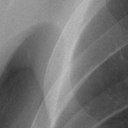}}
\raisebox{-\height}{\includegraphics[width=\textwidth]{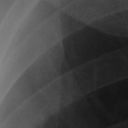}}
\raisebox{-\height}{\includegraphics[width=\textwidth]{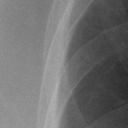}}
 \raisebox{-\height}{\includegraphics[width=\textwidth]{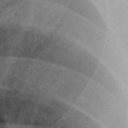}}
 \raisebox{-\height}{\includegraphics[width=\textwidth]{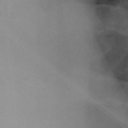}}
 
\caption{Original}
\end{subfigure}

\caption{Comparison of the inpainted versus the real healthy patches. The first column shows the input patch with the region to be inpainted in black. CE: context encoder, SI: semantic inpainting, CA: contextual attention.}
\end{figure}


\begin{table}[ht]
\centering
\begin{tabular}{|l|c|}
  \hline
  Method &  PSNR (mean $\pm$ std) \\
  \hline
  Semantic Inpainting & 33.85 $\pm$ 4.67\\
  \hline
  Context Encoders & 26.31 $\pm$ 4.48 \\
  \hline
  Contextual Attention & 31.80 $\pm$ 5.19  \\
  \hline  
\end{tabular}
\vspace*{0.5em}
\caption{Comparison of PSNR mean and standard deviation on the 880 healthy inpainted patches.}
\label{psnr-healthy-compares}
\end{table}

In contrast, we want to understand how the generative patches would look like if we present the model with abnormal chest x-rays. The ChestXray14 dataset is a comprehensive dataset with many images and sufficient annotations. However, sometimes those annotations are less accurate. To ensure high quality annotations, we randomly choose 33 abnormal chest x-rays labeled as containing lung nodules from the 880 abnormal ones in the test set, and asked an experienced human reader to provide a 64 $\times$ 64 bounding box around a nodule in each x-ray. These 33 x-rays contain a variety of nodular abnormalities. We then add the neighboring 32 pixels on each side of this bounding box to provide some context. Now we have 33 image patches of size 128 $\times$ 128, where each central 64 $\times$ 64 patch contains the abnormality. Then we apply the three generative models to these 33 patches, and the results are demonstrated in Figure 2. The layout of Figure 2 is the same as Figure 1, and we show the average PSNR values for the 33 abnormal regions in Table \ref{psnr-abnormal-compares}. 

As expected, we obtain much less agreement between the generated patches and the real ones, because our generative models have been trained solely on patches extracted from normal images (according to the ChestXray14 labels) so we can expect them to generate healthy looking patches. However, if part of the abnormality shows up outside the mask, such as the third example in Figure 2, then some generative models can produce wrong results by providing continuity with respect to the surrounding abnormal structures. We conclude in this case, the context encoder and the contextual attention model produce better results than the semantic inpainting model.


\begin{figure}[t]
\centering
\begin{subfigure}[t]{0.18\textwidth}
\raisebox{-\height}{\includegraphics[width=\textwidth]{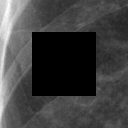}}
\raisebox{-\height}{\includegraphics[width=\textwidth]{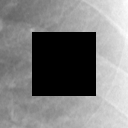}}
\raisebox{-\height}{\includegraphics[width=\textwidth]{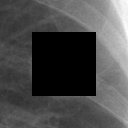}}
\raisebox{-\height}{\includegraphics[width=\textwidth]{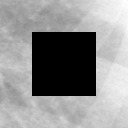}}
\raisebox{-\height}{\includegraphics[width=\textwidth]{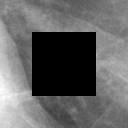}}

\caption{Input}
\end{subfigure}
\begin{subfigure}[t]{0.18\textwidth}
\raisebox{-\height}{\includegraphics[width=\textwidth]{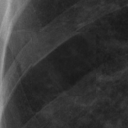}}
\raisebox{-\height}{\includegraphics[width=\textwidth]{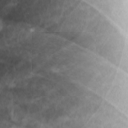}}
\raisebox{-\height}{\includegraphics[width=\textwidth]{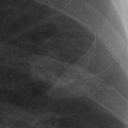}}
\raisebox{-\height}{\includegraphics[width=\textwidth]{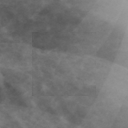}}
\raisebox{-\height}{\includegraphics[width=\textwidth]{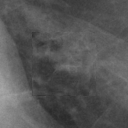}}

\caption{CE}
\end{subfigure}
\begin{subfigure}[t]{0.18\textwidth}
\raisebox{-\height}{\includegraphics[width=\textwidth]{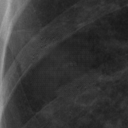}}
\raisebox{-\height}{\includegraphics[width=\textwidth]{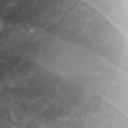}}
\raisebox{-\height}{\includegraphics[width=\textwidth]{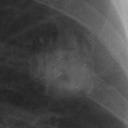}}
\raisebox{-\height}{\includegraphics[width=\textwidth]{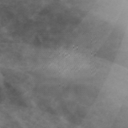}}
\raisebox{-\height}{\includegraphics[width=\textwidth]{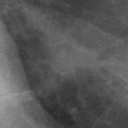}}

\caption{SI}
\end{subfigure}
\begin{subfigure}[t]{0.18\textwidth}   
\raisebox{-\height}{\includegraphics[width=\textwidth]{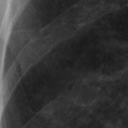}}
\raisebox{-\height}{\includegraphics[width=\textwidth]{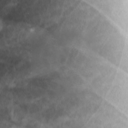}}
\raisebox{-\height}{\includegraphics[width=\textwidth]{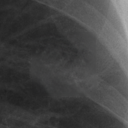}}
 \raisebox{-\height}{\includegraphics[width=\textwidth]{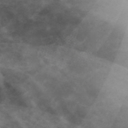}}
  \raisebox{-\height}{\includegraphics[width=\textwidth]{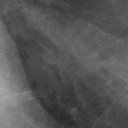}}

 
\caption{CA}
\end{subfigure}
\begin{subfigure}[t]{0.18\textwidth}
\raisebox{-\height}{\includegraphics[width=\textwidth]{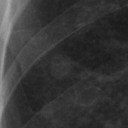}}
\raisebox{-\height}{\includegraphics[width=\textwidth]{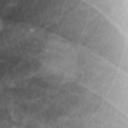}}
\raisebox{-\height}{\includegraphics[width=\textwidth]{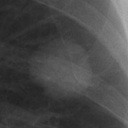}}
\raisebox{-\height}{\includegraphics[width=\textwidth]{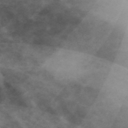}}
\raisebox{-\height}{\includegraphics[width=\textwidth]{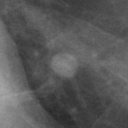}}

\caption{Original}
\end{subfigure}

\caption{Comparison of the generative healthy patches versus the real abnormal patches. CE: context encoder, SI: semantic inpainting, CA: contextual attention.}
\end{figure}

\begin{table}[ht]
\centering
\begin{tabular}{|l|c|}
  \hline
  Method &  PSNR (mean $\pm$ std) \\
  \hline
  Semantic Inpainting & 30.18 $\pm$ 3.28  \\
  \hline
  Context Encoders &  22.22 $\pm$ 4.26 \\
  \hline
  Contextual Attention & 26.79 $\pm$ 3.47 \\
  \hline  
\end{tabular}
\vspace*{0.5em}
\caption{Comparison of PSNR mean and standard deviations on the 33 abnormal patches.}
\label{psnr-abnormal-compares}
\end{table}

Lastly, we randomly select a few healthy and abnormal chest x-ray patches, and subtract our generated patches from the original ones on the pixel-level. As shown in Figure 3 and 4, if the original patch does not have an abnormality, the subtraction produces a region of white noise. This means the differences between the generated and the real ones are small. On the other hand, the subtraction from an abnormal image leaves us a patch with a spherical image structure, which highlights the region of the abnormality. Thus we have demonstrated that the inpainting methods have the potential to be used for enhancement of abnormalities via subtraction.

\begin{figure}[t]

\centering
\begin{subfigure}[t]{0.16\textwidth}
\raisebox{-\height}{\includegraphics[width=\textwidth]{graphs/0_00006249_006_gt.png}}
\raisebox{-\height}{\includegraphics[width=\textwidth]{graphs/0_00011049_002_gt.png}}
\raisebox{-\height}{\includegraphics[width=\textwidth]{graphs/0_00021666_002_gt.png}}
 \raisebox{-\height}{\includegraphics[width=\textwidth]{graphs/0_00020290_001_gt.png}}
 \raisebox{-\height}{\includegraphics[width=\textwidth]{graphs/0_00014447_002_gt.png}}

\caption{Original}
\end{subfigure}
\begin{subfigure}[t]{0.16\textwidth}
\raisebox{-\height}{\includegraphics[width=\textwidth]{graphs/0_00006249_006_generative.png}}
\raisebox{-\height}{\includegraphics[width=\textwidth]{graphs/0_00011049_002_generative.png}}
\raisebox{-\height}{\includegraphics[width=\textwidth]{graphs/0_00021666_002_generative.png}}
 \raisebox{-\height}{\includegraphics[width=\textwidth]{graphs/0_00020290_001_generative.png}}
 \raisebox{-\height}{\includegraphics[width=\textwidth]{graphs/0_00014447_002_generative.png}}
 

\caption{CA}
\end{subfigure}
\begin{subfigure}[t]{0.16\textwidth}   
\raisebox{-\height}{\includegraphics[width=\textwidth]{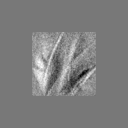}}
\raisebox{-\height}{\includegraphics[width=\textwidth]{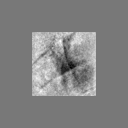}}
\raisebox{-\height}{\includegraphics[width=\textwidth]{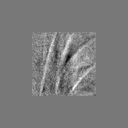}}
 \raisebox{-\height}{\includegraphics[width=\textwidth]{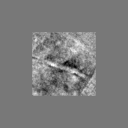}}
 \raisebox{-\height}{\includegraphics[width=\textwidth]{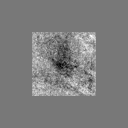}}

 
\caption{Subtraction}
\end{subfigure}
\begin{subfigure}[t]{0.16\textwidth}
\raisebox{-\height}{\includegraphics[width=\textwidth]{00000336_000_real.png}}
\raisebox{-\height}{\includegraphics[width=\textwidth]{00001722_000_gt.png}}
\raisebox{-\height}{\includegraphics[width=\textwidth]{00001322_001_gt.png}}
\raisebox{-\height}{\includegraphics[width=\textwidth]{graphs/00001332_000_gt.png}}
\raisebox{-\height}{\includegraphics[width=\textwidth]{graphs/00000989_000_gt.png}}

\caption{Original}
\end{subfigure}
\begin{subfigure}[t]{0.16\textwidth}
\raisebox{-\height}{\includegraphics[width=\textwidth]{graphs/00000336_000_generative.png}}
\raisebox{-\height}{\includegraphics[width=\textwidth]{00001722_000_generative.png}}
\raisebox{-\height}{\includegraphics[width=\textwidth]{00001322_001_generative.png}}
 \raisebox{-\height}{\includegraphics[width=\textwidth]{graphs/00001332_000_generative.png}}
  \raisebox{-\height}{\includegraphics[width=\textwidth]{graphs/00000989_000_generative.png}}

\caption{CA}
\end{subfigure}
\begin{subfigure}[t]{0.16\textwidth}   
\raisebox{-\height}{\includegraphics[width=\textwidth]{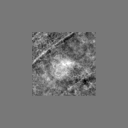}}
\raisebox{-\height}{\includegraphics[width=\textwidth]{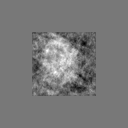}}
\raisebox{-\height}{\includegraphics[width=\textwidth]{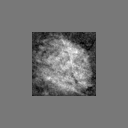}}
 \raisebox{-\height}{\includegraphics[width=\textwidth]{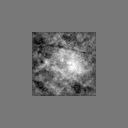}}
  \raisebox{-\height}{\includegraphics[width=\textwidth]{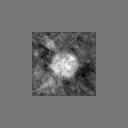}}

 
\caption{Subtraction}
\end{subfigure}


\caption{Subtraction between the original and the inpainted x-rays. Left three columns: normal patches. Right three columns: patches with nodular abnormalities. CA: contextual attention.}
\end{figure}

\subsection{2AFC observer study}
To better quantify the visual quality of the generated patches, we conducted a two-alternative forced choice (2AFC) observer study. We  prepared 160 random pairs of chest x-rays. Both images of the pair are complete 1024 $\times$ 1024 chest x-ray images from the test set, either from the 880 normal or the 880 abnormal images. In one of the two images, a 128 $\times$ 128 patch opverlapping with the segmented lung fields is selected and the central 64 $\times$ 64 region is inpainted by one of the three models in this study. The two images are presented to the observer side by side and the observer has to indicate which of the two images is the real, unaltered, chest x-ray\footnote{The images of this observer study can be found here: \url{https://www.dropbox.com/sh/s2mo1fwuvjvl2mg/AACgg0x_DOl1W6SN8hsesmxsa?preview=0.PNG}}.

The accuracy obtained in these three mixed 2AFC experiments (one for each type of inpainting method) corresponds to the area under the ROC curve \cite{TYLER2000} for distinguishing a real chest x-ray from one where a 64 $\times$ patch overlapping with the lung fields has been inpainted by one of the 3 methods. The observer was a medical researcher with $>20$ years of experience in reading chest radiographs. Table \ref{observe} presents the results of this study. The contextual attention model is the most successful one to fool the expert, because more than half of the time the expert would select the image with an inpainted region as the real chest x-ray. On the other hand, the semantic inpainting model has a more difficult time to fool the expert, where only one third of the time its results are considered `real'. The context encoder's results are usually also not detectable with observer accuracy close to chance levels. These findings are consistent with the results of the previous sections.

\begin{table}[ht]
\centering
\begin{tabular}{|l|c|}
  \hline
  Methods &  Observer accuracy in 2AFC test \\
  \hline
  Semantic Inpainting & 66.66\%   \\
  \hline
  Context Encoders &  59.45\%  \\
  \hline
  Contextual Attention & 37.03 \% \\
  \hline  
\end{tabular}
\vspace*{0.5em}
\caption{Human observer performance for the task of choosing the real x-ray. An accuracy of 50\% corresponds to chance behavior; an accuracy of 100\% indicates the observer can perfectly identify the unaltered chest x-ray in a pair.}
\label{observe}
\end{table}

\section{Discussion and conclusion}

In this paper we applied three modern deep generative models to inpaint chest x-rays. The methods were trained with patches from normal radiographs. We compared the models on their abilities to restore the missing patches in normal chest x-rays, and achieve good visual results. Then we applied these models to abnormal chest x-rays and showed that the results generally produce high contrast between the generated and real image patch which contains the abnormality. For evaluation we used the PSNR metric to assess of the quality of the generated patches in the test set. Not surprisingly, the PSNRs for healthy x-rays are a lot higher than the abnormal ones. In our observer study, we showed that detecting inpainted patches is a difficult task for a human expert, especially when the inpainted regions were generated by the contextual attention model. In future work, we will investigate the performance of inpainting larger regions and subtracting the complete lung inpainted fields, in order to detect subtle abnormalities. We conclude that this work shows for the first time that realistic inpainting in medical images with generative models is feasible.

\subsubsection*{Acknowledgments}
This work was funded by the TTW Perspectief Programme DLMedIA: Deep Learning for Medical Image Analysis (P15-26). We thank the members of DLMedIA for useful discussions. 

\bibliographystyle{unsrt}
\bibliography{refs}

\end{document}